\documentclass[twocolumn]{svjour3}          
\smartqed  
\usepackage{graphicx}
\usepackage{algorithm, longtable}
\usepackage[noend]{algpseudocode}
\usepackage{multirow}
\makeatletter
\def\BState{\State\hskip-\ALG@thistlm}
\makeatother

\usepackage{amsmath,amssymb}

\usepackage{changepage}

\usepackage[utf8x]{inputenc}

\usepackage{textcomp,marvosym}

\usepackage{cite}

\usepackage{nameref,hyperref}

\usepackage[right]{lineno}

\usepackage{microtype}
\DisableLigatures[f]{encoding = *, family = * }

\begin{document}
\title{An Effective Attack Scenario Construction Model based on Attack Steps and Stages Identification
}


\author{Taqwa Ahmed Alhaj\textsuperscript{1*}, Maheyzah Md Siraj\textsuperscript{1}, Anazida Zainal\textsuperscript{1}, Inshirah Idris\textsuperscript{2}, Anjum Nazir\textsuperscript{3},
Fatin Elhaj\textsuperscript{1},Tasneem Darwish\textsuperscript{4}}


\institute{
\textbf{1}Information Assurance and Security Research Group, Faculty of Computing, Universiti Teknologi Malaysia, Johor, Malaysia
\\
\textbf{2} Department of Computer Science and Information Technology, Wad Medani Ahlia University, Wad Medani, Sudan 
\\
\textbf{3} Department of computer Science, Barrett Hodgson University, Karachi, Pakistan
\\
\textbf{4} Department of Systems and Computer Engineering, Carleton University, Ottawa, Canada.
\\
 Corresponding author: 
 *Taqwa-315@hotmail.com
}

\date{Received: date / Accepted: date}

\maketitle
\begin{abstract}
A Network Intrusion Detection System (NIDS) is a network security technology for detecting intruder attacks. However, it produces a great amount of low-level alerts which makes the analysis difficult, especially to construct the attack scenarios. Attack scenario construction (ASC) via Alert Correlation (AC) is important to reveal the strategy of attack in terms of steps and stages that need to be launched to make the attack successful. In most of the existing works, alerts are correlated by classifying the alerts based on the cause-effect relationship. However, the drawback of these works is the identification of false and incomplete correlations due to infiltration of raw alerts. To address this problem, this work proposes an effective ASC model to discover the complete relationship among alerts. The model is successfully experimented using two types of dataset, which are DARPA 2000, and ISCX2012. The Completeness and Soundness of the proposed model are measured to evaluate the overall correlation effectiveness.


\end{abstract}

\section{Introduction}
\label{intro}
The emergence of the Internet of Everything (IoE)
paradigm and the fog/edge computing that brings computing and data analytic very close to the user, requires more advanced security provisioning techniques. Future networks produce and convey large amount of data, and as a result, they are prone to several malicious actions, attacks and security threats that can compromise the availability and integrity of information and services. It is anticipated that future networks will have to ensure the provision of security for a tremendous number of devices with very broad and demanding requirements in an ubiquitous manner.  Therefore, the security and protection of the various communication infrastructures using a Network Intrusion Detection System (NIDS) is of critical importance. A NIDS is a monitoring tool used to monitor and protect networks from attacks. However, a NIDS generates huge amount of low-level intrusion alerts, which makes it difficult to analysis the alerts from these large datasets ~\cite{bib1}~\cite{bib2}. Alert analysis is an essential part for  Security Analyst (SA) task in order to describe the level of significance of an attack. It recognizes the plans or the strategies of intrusions and thereby infers the goal of the attacker~\cite{bib3}. The majority of the research contributions in alert analysis focus on the attack scenario construction to extract attack's intention ~\cite{bib4}. The attack scenario construction elicits the steps and actions taken by the intruder to breach the system. 
An attack scenario is composed of a series of attack stages and each stage contains at least one attack step~\cite{bib3}. An attack step will create several network events. A NIDS determines if a network event can be classified as an intrusion. If the NIDS identifies a network event as an intrusion, then an alert is produced and recorded as shown in Figure 1. 

\begin{figure*}[!h]
\includegraphics[width=0.6\textwidth]{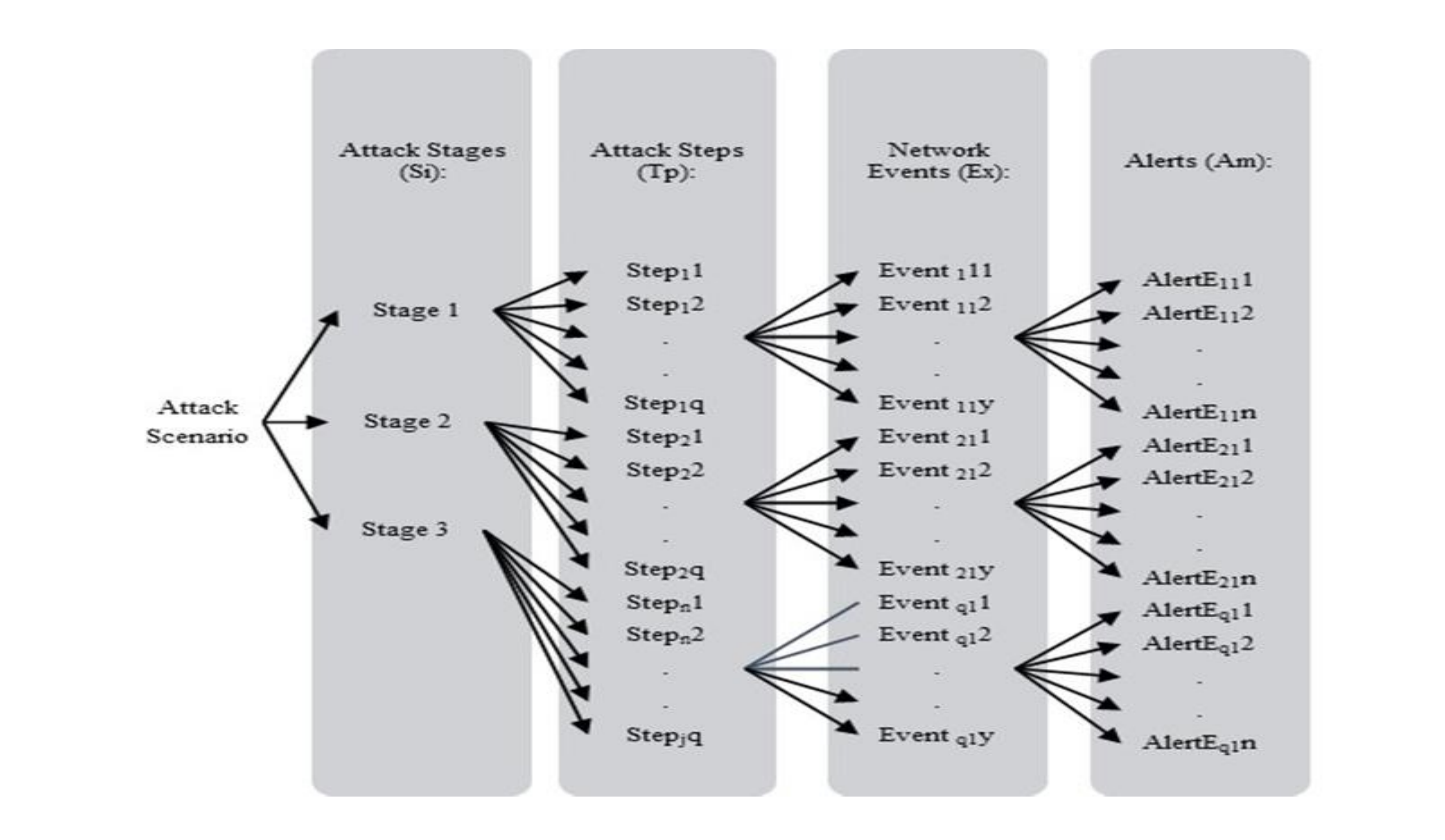}
\caption{\bf The relationship of attack steps and attack stages in attack scenario.}
\label{Fig}
\end{figure*}

Understanding the attack scenario allows the SA to identify the compromised resources, spot the system vulnerabilities, and determine the intruder objectives and the attack severity ~\cite{bib3}. However, the Security Analyst (SA) cannot capture the logical steps or scenarios behind these attacks, due to fact that the NIDS triggers alerts independently in low-level information that describes individual attack steps and are not designed to recognize the attack plans or discover multistage attack scenarios~\cite{bib6}. Therefore, identifying the scenario of the attack directly from these alerts is unmanageable due to problems with detailing a low level of information ~\cite{bib7}~\cite{bib8}. 
Existing works on attack scenario construction mainly either rely on knowledge-based methods to find the relation between alerts or aim to make an inference from statistical or machine learning analysis, which are more complex and higher in computational cost. In addition, in some steps of the attack, the attacker tries many times to compromise the host using different parameters until one attempt succeeds; in this case, the correlation engine generates a redundant relationship between alerts which represents a huge and complex attack scenario. Therefore, this paper investigates method by which to improve the effectiveness of the attack scenario construction. The paper proposed an effective attack scenario construction (ASC) model. The construction of the attack scenario is investigated through three main phases which are: 
\begin{enumerate}
	\item Identifying related alerts: alerts are grouped into inter/intra stages based on the target IP address. Grouping the alerts can help in minimizing the complexity of the attack scenario construction by reducing the number of alerts and improve the ability to describe and visualize the attack scenario. 
	\item Mapping into relevant attack scenario: the main objective of mapping into relevant attack scenario phase is to find candidate hyper alert groupings that are relevant to a particular attack scenario or pattern.
	\item Attack scenario construction: the correlation strength between the alerts inside candidate groups that has complete attack stages is calculated to construct attack scenarios.
\end{enumerate}

The paper begins with some related works in section 2.Then the construction of attack scenario through attack steps and stages is presented in section 3. After that the proposed approach is described in section 4 followed by a detailed investigation on alert grouping based on the target IP ad-dress, alert filtration and attack scenario construction. Results and evaluation analysis are also discussed to show the effectiveness of the proposed model. Finally, a summary of the investigation concludes the paper.

\section{Motivation and related work}
\label{sec:1}

Alert Correlation (AC) is the core of the attack scenario construction process. AC takes a set of alerts produced by one or more NIDSs as input and generates a high-level view of occurring or attempted intrusions \cite{bib8}. It finds and discovers the relationships among unrelated alerts and their attributes that reveal the behavior of the attacker by finding similarity or causality between the alerts \cite{bib3} and \cite{bib12}.
The causality relationship which known as Causal-based Alert Correlation (CAC) \cite{bib13} \cite{bib14} \cite{bib15} \cite{bib9} \cite{bib16} \cite{bib12} and \cite{bib17} tend to find the causality for Attack Scenario Construction (ASC) using three categories, which are Scenario-based; Rule based; and machine learning-based. 
 Scenario-based, some attack scenario template are predefined. Whenever a new alert is received, it is compared with the existing scenarios and then added to the most likely candidate scenario. There are huge numbers of correlation languages related to the specification of attack scenarios have been proposed to implement well-defined scenarios \cite{bib30}. For example, EDL \cite{bib3} is a language for constructing attack scenario. Anew version of EDL is complemented with a method for automatic derivation of multi-step EDL signatures from taint graphs \cite{bib18}. Other researchers used this technique in a complex way by adding additional mechanisms to construct the attack scenario and improve the detection rate. For example, MASP (Mining Attack Sequential Pattern) used incremental mining of subsequences of known multi-step attacks \cite{bib19} and \cite{bib20}. This approach is limited by its need for more comprehensive and scenario complete libraries, and the expenses associated with building and maintaining these libraries.  
Rule-based approaches are one of the main categories used by many researchers \cite{bib21} \cite{bib22} \cite{bib23} \cite{bib24} \cite{bib25} \cite{bib26} \cite{bib27} \cite{bib28} and \cite{bib29}.The knowledge is implemented as conditional, if-then rules. The events when they come are matched with these set of rules \cite{bib30}. Each rule contains two main expressions, which are formulas of predicate calculus linked by an implication connective ($=>$). The left side of the rule contains a prerequisite that must exist for an attack to be finished. The right side, which is consequences, presents the action to be executed if the rule is applicable. For example, [26] proposed a model that represented the strategies of an attack and extracted them from a correlated alert. Their model was described as a directed cyclic graph with nodes that identified attacks and edges that confirmed the temporal order of the attacks. The prerequisites and consequences of known attacks were represented as a hyper-alert type. Lanoe et.al \cite{bib31} developed automata based correlation engine in the context of a European project. A fully automated process generated thousands of correlation rules. Furthermore, in \cite{bib32}, the authors proposed a multi-source fusion model that uses ontologies to represent and store different information resources and reconstructed scenario of the known attack and used a new AOI-FIM algorithm for mining attack patterns of the unknown attack scenarios. This approach does not require profound understanding of the underlying architectural and operational principles of a system. However, it cannot enumerate and encode all possible rules of an individual attack. In addition, the conditions of an attack should not be mistaken for the necessary existence of an earlier attack. 
Machine learning-based employs a different learning algorithm on training dataset and uses knowledge-based data derived from human experts to identify attack scenarios on intrusion patterns and relationships among alerts. Some relation rules or patterns will be created from correlation relationships that satisfy some statistical criteria \cite{bib12}. This involves pair-wise comparisons between alerts since every two alerts might be similar and therefore can be correlated. In this case, the repeated comparisons between alerts will lead to a huge computational overload especially in large scale networks.  Supervised learning algorithms were applied by many authors such as \cite{bib33} \cite{bib34} \cite{bib35} \cite{bib36} \cite{bib37} \cite{bib38} \cite{bib39} \cite{bib40} \cite{bib41} \cite{bib42} and \cite{bib43}. 
Qin and Lee \cite{bib44} proposed an integrated correlation system to identify novel attack strategies using INFOSEC alerts. The authors developed a Bayesian based correlation mechanism to understand and correlate attack steps. Their Bayesian based correlation mechanism employed probabilistic reasoning techniques and included domain knowledge related to individual attacks as a way of understanding and correlating alerts. Furthermore, in the algorithm proposed by \cite{bib36} the level of correlation between nodes in the Bayesian graph depends on the time difference between the events and on the automatic analysis of past events. Bateni et al \cite{bib14} applied correlation matrix in their proposed work. They proposed an alert correlator that consisted of predefined fuzzy rules and dynamic learning-based solutions. Hu et al., \cite{bib45} deployed Absorbing Markov chain (AMC) and big data correlation analysis techniques to estimate the network security. The AMC model was constructed using a large amount of alert data to describe the scenario of multistep attacks. Scalability is one of the advantages in this scheme. However, this method has a high computational cost making it unrealistic for online computation. In addition, in sometimes the attacker tries to compromise the host using different parameters (steps) until one attempt succeeded; in this case, their correlation engine generated a redundant relationship between alerts, which represent huge and complex attack scenario. Furthermore, the previous works construct the attack scenario from raw alerts and do not take into account the sequence and order of attack stages in scenarios, which lead to false and incomplete correlation. Therefore, limitations motivate this paper to construct an effective attack scenario construction (ASC) model.

\section{Attack Scenario Construction through   Steps and Stages}
\label{sec:2}

The attack steps and stages are main components help for ASC. Otherwise, false and incomplete relationship between alerts would be generated \cite{bib9}. Figure 2 shows conceptual relationship among the building blocks investigated in the design and development of the ASC Model. Module labeled as “attack steps” and “attack stages” were discussed in details in our previous work \cite{bib10} \cite{bib11} respectively . The third component is focused on investigation towards the development of ASC model.

In Figure 1, the set of j attack stages in a multi-stage network is represented by Si= {S1, S2, …, Si, ..., Sj}. Each Si is composed of q attack steps that reflect the goals of the attacker. Tp, where p = 1, 2, ..., q and Tp⊆Si, expresses an Attack Step. Every Tp adds to y network events that will be assessed by the NIDS to determine if any intrusive patterns are present. NIDS identified intrusions is expressed as Ex, where x = 1, 2, ..., y and Ex⊆Tp⊆Si. When an Ex occurs, the NIDS will create n alerts to describe the intrusion. Alerts are expressed as Am, where m = 1, 2, ..., n and Am⊆Ex⊆Tp⊆Si. Alert sets are produced and recorded for the SA. SA’s use these unlabeled low-level alerts to examine and understand the attack scenario even though there is no prior knowledge about the underlying cause of the alert. 

\begin{figure}[h]
\includegraphics[width=0.5\textwidth]{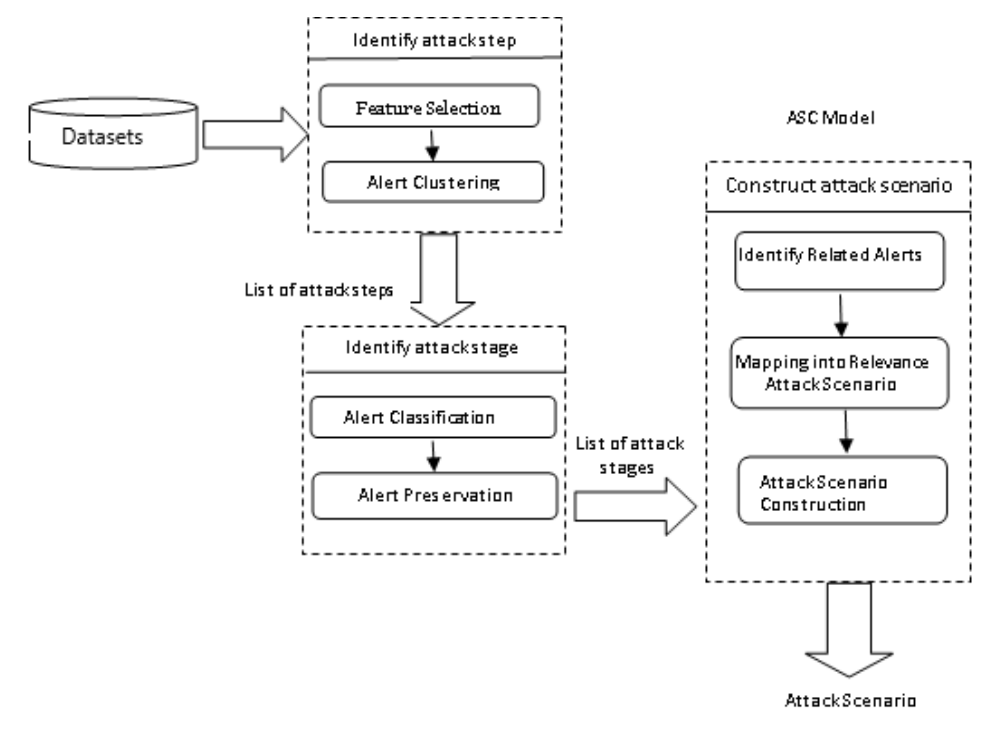}
\caption{\bf : The relationship among components for attack scenario construction}
\label{Fig2}
\end{figure}

\section{Attack Scenario Construction Model}
\label{sec:3}
In this section, an attack scenario construction model for alert correlation is presented as shown in Figure3. The construction of the attack scenario goes through three main processes which are, identify related alert, mapping into relevance attack scenario and calculate correlation strength between alerts.

\begin{figure}[H]
\includegraphics[width=0.45\textwidth]{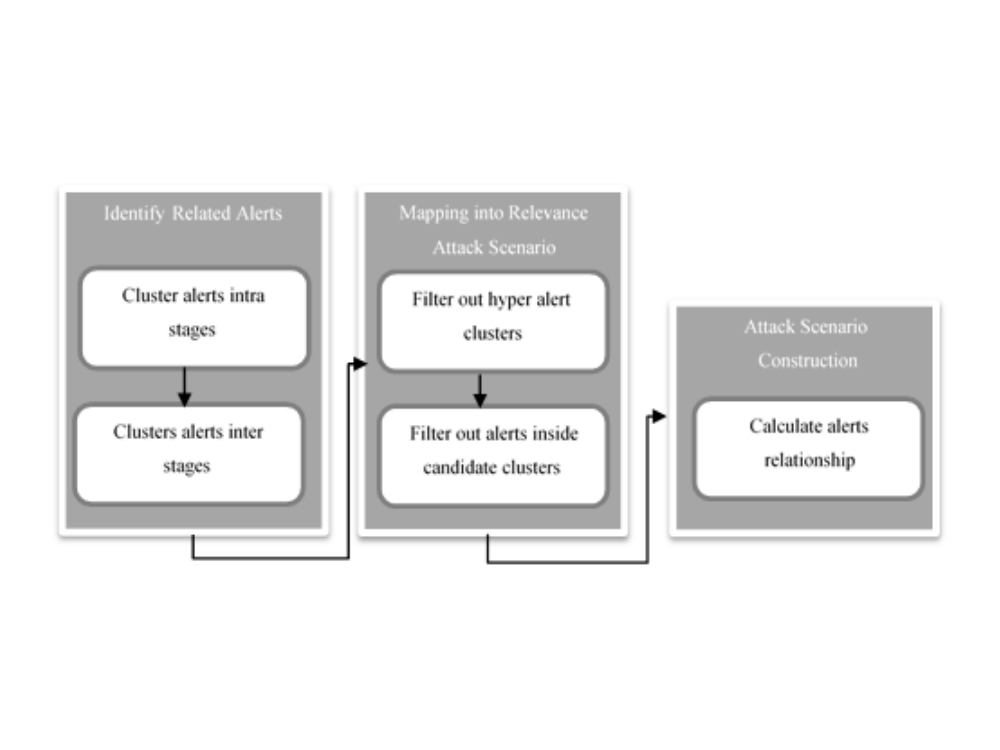}
\caption{\bf The Proposed Attack Scenario Construction Model}
\label{Fig3}
\end{figure}

In identifying related alerts process alerts are grouped inter/intra stages based on the target IP address. While in the mapping process, hyper alert groups that do not contain complete attack stages are filtered out. Finally, the correlation strength between the alerts inside candidate groups that has complete attack stages is calculated to construct attack scenario.

\subsection{Identify Related Alerts}
\label{sec:4}
Identification of the relationships between alerts is an important process to identify related alerts. As mentioned in literature, alert aggregation and clustering can help in minimizing the complexity of the attack scenario construction by reducing the number of alerts and improve the ability to describe and visualize the attack scenario \cite{bib9}. Therefore, alert aggregation and clustering is deployed in this phase to identify related alerts. An earlier study \cite{bib46} mentioned that different attacks in the same attack scenario have the same final intention and most of them are against the same target machine. Thus, the attacks with the same target IP have greater similarity. Therefore, alerts are aggregated into different groups based on their target IPs. Two ways are sequentially used to identify related alerts, which are grouping the alerts into intra stages and grouping the alerts into inter stages.

\begin{enumerate}
	\item Group alerts intra stage: For A set of alerts ${a_1, a_2,…,a_n}$ and S set of stages ${s_1,s_2,…,s_n}$. Given two alerts $(a_1, a_2)\in A$, $a_1$ and $a_2$ can be related into one group the following conditions are met: $a1.target ip = a2.target ip$ and $a_1 ,a_2\in S_i$ 
	where i= 1,2,3,…, The set of alerts A grouped based on the target IP address which occurs when two alerts are grouped together if their target IP addresses are equal. This process is repeated in each of the attack stages S.
	\item Group alerts inter stages: inter stage alerts grouping occurs between stages.	For S set of stages ${s_1,s_2,…,s_n}$ and G set of groups ${g_1,g_2,…,g_n}$. Given two groups $g_1$, $g_2 \in G$ can be related to one group provided the following condition holds:
	          $g_1.target ip =g_2.target ip$  where $g_1 \in S_n$ ,$g_2 \in S_m$ and $n\neq m$
 The output from this grouping is group of alerts from different stages which named as hyper alert groups $A_{hg}$. The proposed algorithm to identify the related alerts is given in Algorithm \ref{Alg1}.
	
\end{enumerate}

\begin{algorithm}
\caption{Identify related alerts}\label{Alg1}
\begin{scriptsize}
\begin{algorithmic}[1]
\Require{$S_i=\{a_1 , a_2, …, a_n\}~~//~alert~instances~in~i^{th}~Stage$}
\Ensure{$Hyper~alert~groups~A_{hg}$}
\State $S_i=~is~a~set~of~attack~stages$
\State $S_1=~1$ 
\State $a_i=~is~alert~instances~in~S_i$
\State $G_i=~is~a~set~of~groups~in~single~stage$
\State $A_{hg}=~is~hyper~alert~groups~among ~different~stages$
\State $// Add~Comments~here$
\For {$i=1: length (a)$}
    \For {$k=i+1: length (a)$}
\If{$(a[S][13].targettip=a[S][15].targettip)$}
\State{$G= [a [S][13].targetip, a [S][15].targetip]$}
\EndIf
\State {\textbf{Endif}}
\EndFor
\State {\textbf{Endfor}}
\EndFor
\State {\textbf{Endfor}}
\For {$j=1: length (G) $}
    \For {$k=1: length (G)$}
\If{$(g[S][14].targetip=g[S+1][15].targetip)$}
\State{$A_{hg} = [g[S][14].targetip, g[S+1][15].targetip]$}
\EndIf
\State {\textbf{Endif}}
\EndFor
\State {\textbf{Endfor}}
\EndFor
\State {\textbf{Endfor}}
\State{$S = S +1$}
\end{algorithmic}
\end{scriptsize}
\end{algorithm}

\subsection{Mapping into Relevant Attack Scenario}
\label{sec:5}

he main objective of this phase is to find candidate hyper alert groups that are relevant to a particular attack scenario or pattern. In order to identify a candidate attack scenario, hyper alert groups that do not contain complete stages and will not contribute in the construction of an attack scenario were filtered out. This phase is divided into two steps, namely: filter out hyper alert groups; and filter out alerts inside candidate hyper alert groups:
\begin{enumerate}
	\item Filter out hyper alert groups: from the result of identifying the related alerts, those hyper alert groups $A_{hg}$ that do not contain alerts belonging to the complete attack stages are filtered out. $	C_{hg} = \phi $ indicates that $A_{hg}$ will not contribute to the construction of the current attack scenario which maybe false positive alerts. Therefore, the outcomes from this step are candidate hyper alerts groups $C_{hg}$ that contain alerts in complete attack stages as specified by formula (3).. For $S_i$ set of stages ${s_1,s_2,…,s_n}$ and $A_{hg}$ set of hyper alert groups ${a_{hg_1},a_{hg_2},…,a_{hg_n}}$

\begin{equation}
    	C_{hg}= \begin{Bmatrix} \forall a_{hg} \in  A_{hg} \exists a_{hg}\in S_{i}\forall i=1,2,...,n	\\ \phi ,  Otherwise\end{Bmatrix} 
\end{equation}

	\item Filter out alerts inside candidate groups: Referring to candidate hyper alerts groups 	$C_{hg}$, the timestamp of each of the alerts in this group is taken into account. The timestamp of the alerts that are classified in the earlier attack stages is compared to the timestamp of the alerts in the last attack stage denoting as: 
\begin{scriptsize}
\begin{equation}
	C_{hg}= \begin{Bmatrix} \forall a_{i} \in C_{hg}, a_T.S_{n-1} \le a_T.S_n \forall i=1,2,...,n	\\ \phi ,  Otherwise\end{Bmatrix} 
\end{equation}	
\end{scriptsize}
	An observation has been noted suggestion that there are some alerts $a_i\in C_{hg}$ have greater timestamp than the timestamp of alerts in the last attack stage. Based on this observation, the alert filtration phase filters out and discards those alerts. These indicate that those alerts that are triggered after the last stage of an attack can present another attack attempt. The detailed proposed algorithm that demonstrates the mapping of alerts into relevant attack scenario is shown in Algorithm \ref{Alg2} 
\end{enumerate}

\begin{algorithm}
\caption{Mapping into relevant attack scenario }\label{Alg2}
\begin{scriptsize}
\begin{algorithmic}[1]
\Require{$S_i=\{a_1 , a_2, …, a_n\}~~//~alert~instances~in~i^{th}~Stage$}
\Ensure{$Candidate~hyper~alert~groups~C_{hg}$}
\State $S_i=~is~a~set~of~attack~stages$
\State $a_i=~is~alert~instances~in~S_i$
\State $C_{hg}[~]=~is~candidate~hyper~alert~group$
\State $A_{hg}=~is~hyper~alert~groups~among ~different~stages$
\State $// Add~Comments~here$
\For {$i=1: length (S_i)$}
    \For {$j=1: length (A_{hg})$}
    \State{$C_{hg}= A_{hg}~[13][14]$}
\EndFor
\State {\textbf{Endfor}}
\EndFor
\State {\textbf{Endfor}}
\If{$((A_{hg} = =C_{hg}))$}
\State{$C_{hg}= A_{hg}$}
\Else
\State{$Delete~A_{hg}$}
\EndIf
\State {\textbf{Endif}}
\State{$Last_{stage}=length(C_{hg})$}
\State{$ T = length (Last_{stage})).time$}
\For {$i=1: length (C_{hg})$}
    \For {$i=j: length (S_i)$}
        \If{$If (S_i[13][14]).time < T$}
            \State{$C_{hg}= a[13][14]$}
        \Else
            \State{$Delete~C_{hg}$}
        \EndIf
    \EndFor
    \State {\textbf{Endfor}}   
\EndFor
\State {\textbf{Endfor}}
\end{algorithmic}
\end{scriptsize}
\end{algorithm}

\subsection{Attack Scenario Construction}
\label{sec:6}
A hyper-alert graph $HG = (N, E)$ is a connected DAG 
(Directed A cyclic Graph) where the set N of nodes is a set of alerts inside the candidate group. For each pair of nodes $n_1, n_2 \in N$, there is an edge from $n_1$ to  $n_2$ in E. This is if and only if $n_1$ belongs to a previous attack stage, $n_2$ belongs to the next attack stage and there is correlation strength between $n_1$ and $n_2$.
The correlation strength between two types of alerts plays an important role in attack pattern analysis as it reveals the relationship between two alerts. Linear correlation coefficient is used to measure the strength and the direction of a linear relationship between two alerts. It is sometimes referred to as the Pearson correlation coefficient. Consider that two types of alerts $a_1$ and $a_2$ are given; to calculate the correlation strength r between them, the following mathematical formula is applied: 

\begin{equation}
r=\frac{n\sum{a_1 a_2 -(\sum a_1)(\sum a_2)}}{\sqrt{n(\sum a_1^2)}-(\sum a_1)^2\sqrt{n(\sum a_2^2)}-(\sum a_2)^2)}
\end{equation}

The value of $r$ is such that $-1<r<+1 $. The $+$ and $–$ signs are used for positive linear correlations and negative linear correlations, respectively.  Positive correlation is considered as follows: If $a_1$ and $a_2$ have a strong positive linear correlation, $r$ is close to $+1$. Positive values indicate a relationship between $a_1$ and $a_2$ such that as values for $a_1$ increase, values for $a_2$ also increase. Conversely, with negative correlation, if $a_1$ and $a_2$ have a strong negative linear correlation, $r$ is close to $-1$.

Negative values indicate a relationship between $a_1$ and $a_2$ such that as values for $a_1$ increase, values for $a_2$ decrease. If there is no linear correlation or a weak linear correlation, $r$ is close to $0$. A value near zero means that there is a random, nonlinear relationship between the two variables.The detailed proposed algorithm  is presented in Algorithm \ref{Alg3}.Source IP and Target IP are removed from alert features when calculating the correlation strength between alerts, since there is a concern that the attackers may spoof the source IP address. The target IP is the same as we have encountered in all alerts.

\begin{algorithm}
\caption{Attack Scenario Construction}\label{Alg3}
\begin{scriptsize}
\begin{algorithmic}[1]
\Require{$S_i=\{a_1 , a_2, …, a_n\}~~//~alert~instances~in~i^{th}~Stage$}
\Ensure{$Attack~scenario~graph,~Completeness,~Soundness$}

\State $C_{final}=~is~a~final~set~of~alerts~inside~the~candidate~group$
\State $Construct[~]=~correlation~strength~between~two~of~alerts~$
$inside~candidate~group$
\State $Graph_{i}=~is~the~generated~graph~number$
\State $CM~is~a~variable~for~measuring~completeness$
\State $SU~is~a~variable~for~measuring~Soundness$
\For {$a_{i}: length(C_{final})$}
    \State{$Construct[~]=Corr(a_i , a_{i+1})$}
    \State{$Graph_{i}= Draw(Construct[~])$}
    \State{$CM =Completeness (Graph_{i})$}
    \State{$SU =Soundness (Graph_{i})$}        
\EndFor
\State {\textbf{Endfor}}
\end{algorithmic}
\end{scriptsize}
\end{algorithm}

\section{Experiments Result and Discussion}
\label{sec:7}

To evaluate the effectiveness of the proposed model in constructing effective attack scenarios, the experiments were conducted on two different datasets which are DARPA 2000 (inside network dataset in scenarios LLDOS 1.0 and LLDOS 2.0.2) and ISCX 2012 dataset. 
In DARPA 2000, these attack scenarios are carried out over multiple network and audit sessions. These sessions have been grouped into five (5) attack stages over the course of which the adversary: probes; breaks-in; installs trojan Mstream DDoS software; and launches a DDoS attack against an off-site server. As a part of this process, the attacker uses the Solaris sadmind exploit, a well-known Remote-To-Root attack. This successfully gained root access to three Solaris hosts which are Mill (172.16.115.20), Pascal (172.16.112.50), and Locke (172.16.112.10) at Eyrie Air Force Base (AFB) Network. Meanwhile in ISCX 2012 dataset a full attack scenario is contained five attack. The attack scenario was launched on 21 interconnected workstations that were running windows software. The interconnected network composed of four (4) distinct LANS. One LAN provided web, email, DNS, and Network Address Translation (NAT) services. The NAT server (192.168.5.124) acted as an internet access provider for the entire network. Two valid IP addresses connected the NAT server the internet. The main server had one of these IP addresses dedicated to it, while the other was multiplexed across every workstation.  The networks website, email, and internal name resolvers were handled by the main server (192.168.5.122). 

\subsection{Result for Identifying Related Alerts}
\label{sec:8}
This section presents the result of identifying a related alerts phase that discussed in previous section. At the end of this phase, 33 and 27 groups of alerts with different numbers of attack stages were obtained from LLDOS 1.0 and LLDOS 2.0.2 respectively. In ISCX 2012 dataset, the total of 363 hyper alerts groups are obtained, 342 groups of alerts from the total groups belong to a single attack stage while 21 groups were obtained with different numbers of attack stages.
Figures 4,5 and 6 present the hyper alert groups in LLDOS 1.0, LLDOS 2.0.2 and ISCX that belong to different number of attack stages. Five candidates’ hyper alerts groups were identified and selected as related alerts in all datasets. In LLDOS 1.0 scenario, the candidate hyper alerts groups are based on 172.016.112.010, and 172.016.112.050, which contain complete attack stages. In LLDOS2.0.2 dataset, the candidate hyper alert groups are based on 172.016.115.020 and contain complete attack stages. Based on the description of the dataset, the attack script has built a list of those hosts on which it has success-fully installed the mstream zombie and launched the DDoS. These are mill (172.16.115.20), pascal (172.16.112.50), and locke (172.16.112.10) respectively. Therefore, our model allows for the successful finding of related alerts that contribute to the construction of an effective attack scenario. 
While, in ISCX 2012 dataset, the hyper alerts groups that belong to complete attack stages are based on 192.168.5.122 and 192.168.2.107. Referring to the dataset description found that the first session of initiated attack starting by scanning potential target hosts on two consecutive subnets (192.168.1.0/24 and 192.168.2.0/24). A running Windows XP SP1 with vulnerable SMB authentication protocol is identified. This vulnerability is exploited and a scan is performed to the server subnet (192.168.5.0/24). As result of this finding the proposed model is successfully identify the related alerts that will contribute in the construction of the model.
\begin{figure}[h]
\includegraphics[width=0.4\textwidth]{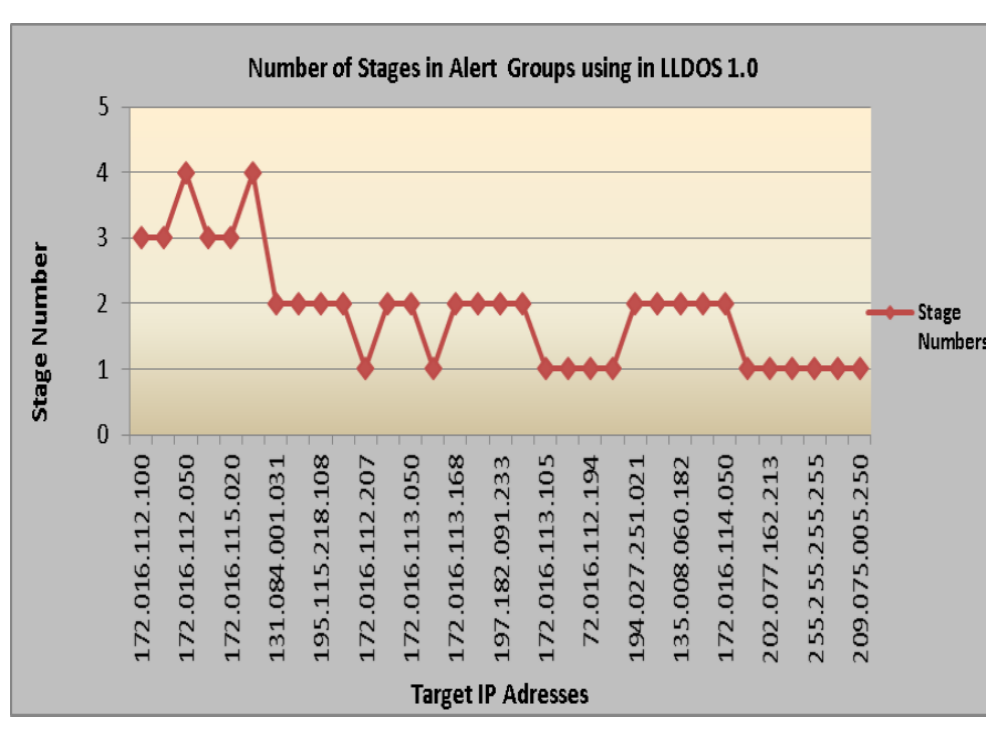}
\caption{\bf Number of attack stages in alert groups based on target IP addresses in LLDOS1.0}
\label{Fig4}
\end{figure}
\begin{figure}[h]
\includegraphics[width=0.4\textwidth]{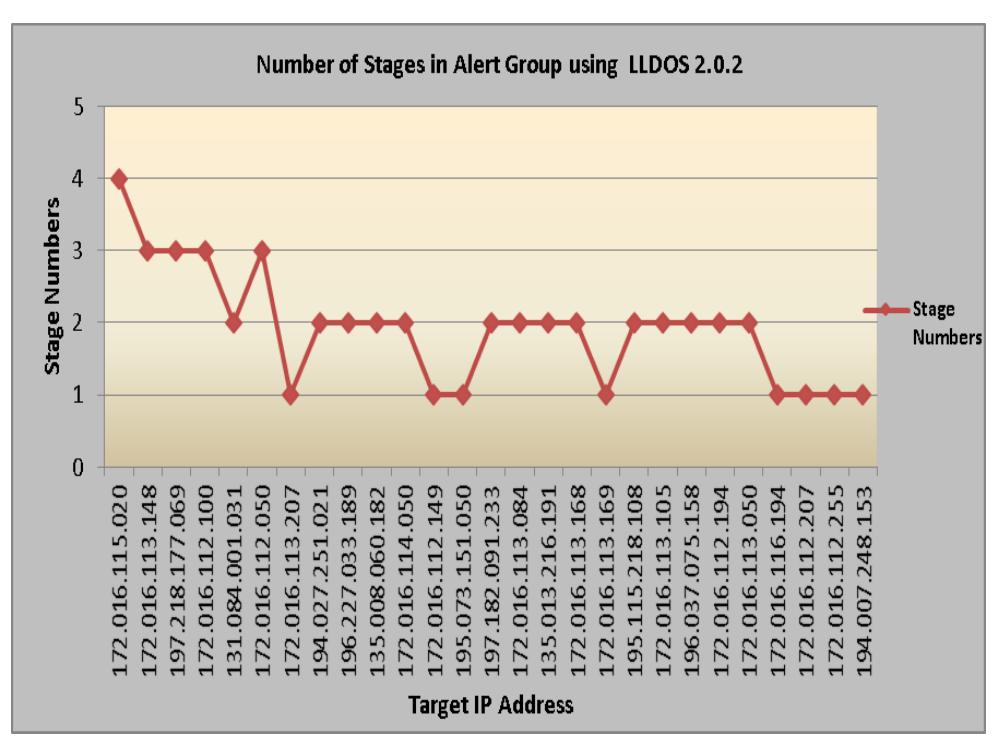}
\caption{\bf Number of attack stages in alert groups based on target IP addresses in LLDOS 2.0.2}
\label{Fig5}
\end{figure}
\begin{figure}[h!]
\includegraphics[width=0.4\textwidth]{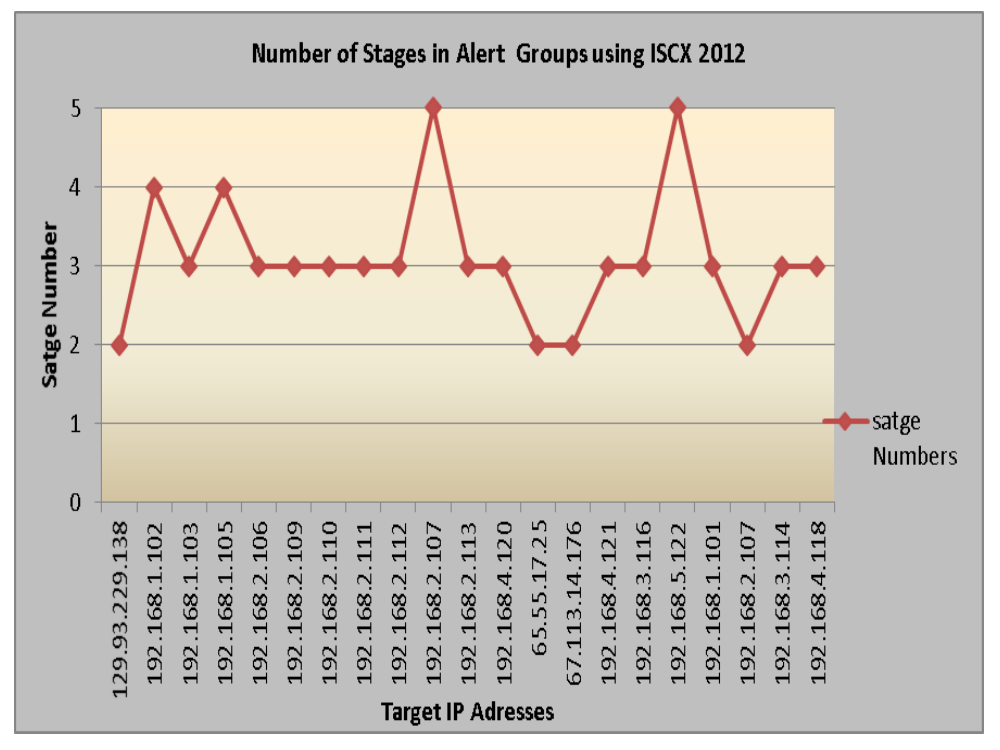}
\caption{\bf Number of attack stages in alert groups based on target IP addresses in ISCX2012 }
\label{Fig6}
\end{figure}

\subsection{Result on Mapping into Relevant Attack Scenario}
\label{sec:9}
This phase maps the hyper alert groups into the relevant attack scenario by filtering out the hyper alert groups that do not contain complete attack stages. From identifying related alerts, results show that, out of the total number of hyper alert groups (33, 27 and 363) in all datasets there are five hyper alerts groups related to the relevant attack scenario. Out of the remaining 31, 26 and 361 groups of alerts are filtered out because they may launch single attack attempts that are irrelevant to any attack scenario. The amount of alerts in the three candidate hyper alerts groups that have been used in attack scenario construction are given in Tables 1, 2 and 3. Furthermore, this phase discards those alerts that have a greater time stamp than the time stamp of alerts in the last attack stage. Tables 4,5 and 6 show the final amount of alerts that will contribute in the attack scenario construction inside five candidate groups for each of the datasets.

\begin{table*}[]
\caption{Amount of alerts inside the candidate hyper alert groups in LLDOS 1.0}
\label{Table1}
\begin{tabular}{|p{0.1\linewidth}|l|l|p{0.2\linewidth}|}
\hline
\multicolumn{1}{|l|}{Target IP addresses} & Alert Type   & Amount & Attack Stages                                \\ \hline
\multirow{12}{*}{172.016.112.050}                      & FTP\_Syst                     & 2      & \multirow{2}{=}{Pre-attack Probe}            \\ \cline{2-3}
                                                       & Sadmind\_Ping                 & 1      &                                              \\ \cline{2-4} 
                                                       & Admind                        & 5      & \multirow{4}{=}{Unauthorized Access Attempt} \\ \cline{2-3}
                                                       & Email\_Ehlo                   & 17     &                                              \\ \cline{2-3}
                                                       & Email\_Almail\_Overflow       & 2      &                                              \\ \cline{2-3}
                                                       & Sadmind\_Amslverify\_Overflow & 4      &                                              \\ \cline{2-4} 
                                                       & FTP\_Pass                     & 3      & \multirow{5}{=}{Protocol Signature}          \\ \cline{2-3}
                                                       & FTP\_User                     & 3      &                                              \\ \cline{2-3}
                                                       & Rsh                           & 1      &                                              \\ \cline{2-3}
                                                       & SSH\_Detected,                & 2      &                                              \\ \cline{2-3}
                                                       & TelnetTerminaltype            & 31     &                                              \\ \cline{2-4} 
                                                       & Mstream\_Zombie               & 1      & Suspicious Activity                          \\ \hline
\multicolumn{1}{|l|}{\multirow{5}{*}{172.016.112.010}} & Sadmind\_Ping                 & 1      & Pre-attack Probe                             \\ \cline{2-4} 
\multicolumn{1}{|l|}{}                                 & Sadmind\_Amslverify\_Overflow & 4      & \multirow{2}{=}{Unauthorized Access Attempt} \\ \cline{2-3}
\multicolumn{1}{|l|}{}                                 & Admind                        & 5      &                                              \\ \cline{2-4} 
\multicolumn{1}{|l|}{}                                 & Rsh                           & 1      & Protocol Signature                           \\ \cline{2-4} 
\multicolumn{1}{|l|}{}                                 & Mstream\_Zombie               & 1      & Suspicious Activity                          \\ \hline
\end{tabular}
\end{table*}

\begin{table*}[]
\caption{Amount of Alerts inside the candidate hyper alert groups in LLDOS 2.0.2}
\label{Table2}
\begin{tabular}{|p{0.2\linewidth}|p{0.3\linewidth}|p{0.1\linewidth}|p{0.2\linewidth}|}
\hline
Target IP addresses & Alert Types & Amount & Attack Stages \\ \hline
\multirow{10}{=}{172.016.115.020} & FTP\_Syst & 1 & Pre-attack Probe \\ \cline{2-4} 
 & Admind & 2 & \multirow{2}{=}{Unauthorized Access Attempt} \\ \cline{2-3}
 & Sadmind\_Amslverify\_ Overflow & 2 &  \\ \cline{2-4} 
 & FTP\_Pass & 2 & \multirow{6}{=}{Protocol Signature} \\ \cline{2-3}
 & FTP\_Put & 1 &  \\ \cline{2-3}
 & FTP\_User & 2 &  \\ \cline{2-3}
 & Telnet Terminaltype & 4 &  \\ \cline{2-3}
 & TelnetEnvAll & 1 &  \\ \cline{2-3}
 & TelnetXdisplay & 1 &  \\ \cline{2-4} 
 & Mstream\_Zombie & 2 & Suspicious Activity \\ \hline
\end{tabular}
\end{table*}

\begin{table*}[]
\caption{Amount of Alerts inside the candidate hyper alert groups in ISCX 2012}
\label{Table3}
\begin{tabular}{|p{0.2\linewidth}|p{0.3\linewidth}|p{0.1\linewidth}|p{0.2\linewidth}|}
\hline
Target IP addresses & Alert Type & Amount & Attack Stages \\ \hline
\multirow{8}{=}{192.168.5.122} & ICMP test & 103 & Information gathering \\ \cline{2-4} 
 & PROTOCOL-DNS potential dns cache poisoning attempt & 7622 & \multirow{3}{=}{Vulnerability identification and scanning} \\ \cline{2-3}
 & ET SCAN Potential SSH Scan & 2 &  \\ \cline{2-3}
 & ET SCAN LibSSH Based Frequent SSH Connections & 8 &  \\ \cline{2-4} 
 & PROTOCOL-DNS TMG Firewall Client entry exploit attempt & 7444 & \multirow{2}{=}{Gaining access and compromising a system} \\ \cline{2-3}
 & FTP\_TELNETET & 3526 &  \\ \cline{2-4} 
 & INFO SUSPICIOUS SMTP EXE - EXE SMTP Attachment & 2 & Maintaining access and installing suspicious behaviour \\ \cline{2-4} 
 & (http\_inspect) & 1 & Covering Attack \\ \hline
\multirow{4}{=}{192.168.2.107} & ICMP test & 3837 & Information gathering \\ \cline{2-4} 
 & ET POLICY PE EXE or DLL Windows file download HTTP & 1 & Gaining access and compromising a system \\ \cline{2-4} 
 & ET INFO Executable Retrieved With Minimal HTTP Headers & 2 & Maintaining access and installing suspicious behaviour \\ \cline{2-4} 
 & (spp\_frag3) Tiny fragment & 1 & Covering Attack \\ \hline
\end{tabular}
\end{table*}

\begin{table*}[]
\caption{Total amount of alerts used in the construction of attack scenario in LLDOS 1.0}
\label{Table4}
\begin{tabular}{|p{0.2\linewidth}|p{0.3\linewidth}|p{0.1\linewidth}|p{0.2\linewidth}|}
\hline
Target IP addresses & Alert Types & Amount & Attack Stages \\ \hline
\multirow{12}{=}{172.016.112.050} & FTP\_Syst & 1 & \multirow{2}{=}{Pre-attack Probe} \\ \cline{2-3}
 & Sadmind\_Ping & 1 &  \\ \cline{2-4} 
 & Admind & 5 & \multirow{4}{=}{Unauthorized Access Attempt} \\ \cline{2-3}
 & Email\_Ehlo & 10 &  \\ \cline{2-3}
 & Email\_Almail\_Overflow & 2 &  \\ \cline{2-3}
 & Sadmind\_Amslverify\_ Overflow & 4 &  \\ \cline{2-4} 
 & FTP\_Pass & 2 & \multirow{5}{=}{Protocol Signature} \\ \cline{2-3}
 & FTP\_User & 2 &  \\ \cline{2-3}
 & Rsh & 1 &  \\ \cline{2-3}
 & SSH\_Detected & 1 &  \\ \cline{2-3}
 & TelnetTerminaltype & 23 &  \\ \cline{2-4} 
 & Mstream\_Zombie & 1 & Suspicious Activity \\ \hline
\multirow{5}{=}{172.016.112.010} & Sadmind\_Ping & 1 & Pre-attack Probe \\ \cline{2-4} 
 & Sadmind\_Amslverify\_ Overflow & 4 & \multirow{2}{=}{Unauthorized Access Attemp} \\ \cline{2-3}
 & Admind & 5 &  \\ \cline{2-4} 
 & Rsh & 1 & Protocol Signature \\ \cline{2-4} 
 & Mstream\_Zombie & 1 & Suspicious Activity \\ \hline
\end{tabular}
\end{table*}

\begin{table*}[]
\caption{Total amount of alerts used in the construction of attack scenario in LLDOS 2.0.2}
\label{Table5}
\begin{tabular}{|p{0.2\linewidth}|p{0.3\linewidth}|p{0.1\linewidth}|p{0.2\linewidth}|}
\hline
Target IP addresses & Alert Types & Amount & Attack Stages \\ \hline
\multirow{9}{=}{172.016.115.020} & Admind & 2 & \multirow{2}{=}{Unauthorized Access Attempt} \\ \cline{2-3}
 & Sadmind\_Amslverify\_ Overflow & 2 &  \\ \cline{2-4} 
 & FTP\_Pass & 1 & \multirow{6}{=}{Protocol Signature} \\ \cline{2-3}
 & FTP\_Put & 1 &  \\ \cline{2-3}
 & FTP\_User & 1 &  \\ \cline{2-3}
 & Telnet Terminaltype & 3 &  \\ \cline{2-3}
 & TelnetEnvAll & 1 &  \\ \cline{2-3}
 & TelnetXdisplay & 1 &  \\ \cline{2-4} 
 & Mstream\_Zombie & 2 & Suspicious Activity \\ \hline
\end{tabular}
\end{table*}

\begin{table*}[]
\caption{Total amount of alerts used in the construction of attack scenario in ISCX2012}
\label{Table6}
\begin{tabular}{|p{0.2\linewidth}|p{0.3\linewidth}|p{0.1\linewidth}|p{0.2\linewidth}|}
\hline
Target IP addresses & Alert Type & Amount & Attack Stages \\ \hline
\multirow{5}{=}{192.168.5.122} & ICMP test & 3 & Information gathering \\ \cline{2-4} 
 & PROTOCOL-DNS potential dns cache poisoning attempt & 1356 & Vulnerability identification and scanning \\ \cline{2-4} 
 & PROTOCOL-DNS TMG Firewall Client entry exploit attempt & 1333 & Gaining access and compromising a system \\ \cline{2-4} 
 & ET INFO SUSPICIOUS SMTP EXE - EXE SMTP Attacment & 1 & Maintaining access and installing suspicious behaviour \\ \cline{2-4} 
 & (http\_inspect) & 1 & Covering Attack \\ \hline
\multirow{4}{=}{192.168.2.107} & ICMP test & 3356 & Information gathering \\ \cline{2-4} 
 & ET POLICY PE EXE or DLL Windows file download HTTP & 1 & Gaining access and compromising a system \\ \cline{2-4} 
 & ET INFO Executable Retrieved With Minimal HTTP Headers & 2 & Maintaining access and installing suspicious behaviour \\ \cline{2-4} 
 & (spp\_frag3) Tiny fragment & 1 & Covering Attack \\ \hline
\end{tabular}
\end{table*}

\subsection{Result on Attack Scenario Construction}
\label{sec:10}

In order to construct the attack scenario, the correlation strength between two types of alerts should be measured in each list of alerts inside the candidate hyper alert groups that come out from the alert filtration phase. The details on the correlation strength of alerts in each candidate hyper alert groups for all dataset are given in Tables 7, 8, 9, 10 and 11 respectively. Note that in each of the candidate groups, there are repeated alerts, which represent the situation so that the attacker may repeat these steps in order to succeed. Those repeated alerts are represented as one type of alert in the correlation strength tables and as self-looped cycles in the graph.

Five graphs are generated by the correlation strength representing the scenarios in which the attacker uses the same stages to compromise the target hosts in the network as shown in Figures 7, 8, 9, 10 and 11.

In DARPA 2000 (LLDOS1.0 and LLDOS2.0.2) there are three attack scenarios. The LLDOS1.0 has two attack scenarios that compromise locke (172.16.112.10) and pascal (172.16.112.50) network hosts. Meanwhile, LLDOS2.0.2 has one attack scenario, which compromised mill (172.16.115.20) host. The attack scenario which compromised the locke host consistent with the description in the DARPA LLDOS1.0 documentation. In addition to this scenario, our proposed model identifies other scenarios relating to compromising the pascal (172.16.112.50) host. For example, an attacker tries to gain unauthorized access to several hosts inside the network. This activity triggers multiple Email Ehlo and Email\_Almail\_Overflow alerts. After gaining root access for the host, the attacker uses Ftp and Telnet to install mstream Trajon causing a trigger for multiple FTP alerts such as FTP User, FTP Pass, FTP Syst, TelnetTerminaltype, TelnetEnvAll and TelnetXdisplay. LLDOS 2.0.2 attack scenario has a similar pattern with the LLDOS1.0 scenario. In both cases, the attacker compromises several machines (in different ways) by exploiting the vulnerability of the Sadmind service and installs DDoS daemons on these machines by using rsh, telnet or ftp. Figure 7 is the corresponding graph that is generated from Table 7 to represent the attack scenario that compromised locke (172.16.112.10) host in the same dataset. The relation between the alerts in this graph also is very strong, except that the Mstream\_Zombie has weak relations with Sadmind\_Ping and Rsh alerts. Figure 8 presents the attack scenario for compromising pascal (172.16.112.50) host in LLDOS1.0 that was drawn from the corresponding table of correlation strengths of alerts in 172.016.112.050 group. It shows the major steps and all the possible transitions that the attack followed in order to achieve its goal. In addition, it demonstrates that most of the alerts have a strong correlation with each other. The exception to this is the SSH alert which has a weak correlation in some alerts (as shown in bold font in Table 8). Figure 9 is the attack graph extracted from Table 9. It is easy to determine that all alerts have a strong correlation with each other.
In ISCX 2012 two attack scenarios are identified by the proposed model intended to compromise two main hosts, 192.168.5.122 and 192.168.2.107. In 192.168.5.122 infected host an attacker, tries to perform DDoS attack using Botnet by combining multiple attack scenarios as described in the dataset documentation. For example, an attacker tries to gather a lot of information about the target host. This activity triggers multiple ICMP test and PROTOCOL\-DNS potential dns alerts. Then an attacker attempts to exploit some vulnerabilities via unauthorized access to several hosts causing a trigger for multiple PROTOCOL\-DNS potential dns alerts. After gaining access for the host, the attacker installed malicious software causing a trigger for ET INFO SUSPICIOUS SMTP EXE\-ATTACHMENT alerts. Finally, The Distributed Denial of Service attack is started using the run http IP threads command causing a trigger for a large number of HTTP\_INSPECT alerts. Figure 10 is the corresponding graph that is generated from Table10 to represent the attack scenario. It demonstrates the major steps and all the possible transitions that the attack followed in order to achieve its target. Additionally, the figures demonstrate that the relation between the alerts in this graph is very strong, except the http\_inspect alert has some weak relations with some alerts. Meanwhile the second attack scenario that compromised 192.168.2.107 host is identified toward performing a separate denial of service attack by performing an HTTP GET attack. An attacker exploits HTTP GET requests to attack the host of 192.168.2.107.  This activity triggers multiple ET POLICY PE EXE download HTTP and ET INFO Executable Retrieved with Minimal HTTP Headers. Figure 11 presents the attack scenario for compromising 192.168.2.107 host that was drawn from the corresponding table of correlation strengths of alerts in 192.168.2.107 group. The Figure easy shows that all the alerts have a strong correlation with each other.

\begin{table*}[]
\caption{The correlation strength of alerts in 172.016.112.010 cluster using LLDOS 1.0}
\label{Table7}
\begin{tabular}{|p{0.15\linewidth}|p{0.15\linewidth}|p{0.1\linewidth}|p{0.15\linewidth}|p{0.1\linewidth}|p{0.15\linewidth}|}
\hline
Alerts & Sadmind\_ Ping & Admind & Sadmind\_ Amslverify\_ Overflow & Rsh & Mstream\_ Zombie \\ \hline
Sadmind\_ Ping & 1 & 0.7182 & 0.6699 & -0.9417 & 0.0474 \\ \hline
Admind & 0.7182 & 1 & 0.9978 & -0.9104 & 0.7291 \\ \hline
Sadmind\_ Amslverify\_ Overflow & 0.6699 & 0.9978 & 1 & -0.8806 & 0.7734 \\ \hline
Rsh & -0.9417 & -0.9104 & -0.8806 & 1 & -0.3806 \\ \hline
Mstream\_ Zombie & 0.0474 & 0.7291 & 0.7734 & -0.3806 & 1 \\ \hline
\end{tabular}
\end{table*}

\begin{table*}[]
\caption{The correlation strength of alerts in 172.016.112.050cluster using LLDOS 1.0}
\label{Table8}
\begin{tiny}
\begin{tabular}{|p{0.06\linewidth}|p{0.04\linewidth}|p{0.06\linewidth}|p{0.05\linewidth}|p{0.04\linewidth}|p{0.05\linewidth}|p{0.06\linewidth}|p{0.04\linewidth}|p{0.04\linewidth}|p{0.04\linewidth}|p{0.04\linewidth}|p{0.05\linewidth}|p{0.06\linewidth}|}
\hline
Alerts & FTP\_ Syst & Sadmind Ping & Admind & Email\_ Ehlo & Email\_ Almail\_ Overflow & Sadmind Amslverify\_ Overflow & FTP\_ Pass & FTP\_ User & Rsh & SSH\_ Detected & Telnet Terminal type & Mstream Zombie \\ \hline
FTP\_ Syst & 1 & 0.7013 & 0.7081 & 1 & 0.9999 & 0.7081 & 1 & 1 & 1 & 0.9621 & 0.9996 & 0.9966 \\ \hline
Sadmind Ping & 0.7013 & 1 & 1 & 0.707 & 0.7119 & 1 & 0.7013 & 0.7013 & 0.6986 & 0.4804 & 0.7207 & 0.7638 \\ \hline
Admind & 0.7081 & 1 & 1 & 0.7138 & 0.7186 & 1 & 0.7081 & 0.7082 & 0.7055 & 0.4889 & 0.7274 & 0.7638 \\ \hline
Email\_ Ehlo & 1 & 0.707 & 0.7138 & 1 & 1 & 0.7138 & 1 & 1 & 0.9999 & 0.9599 & 0.9998 & 0.9972 \\ \hline
Email\_ Almail\_ Overflow & 0.9999 & 0.7119 & 0.7186 & 1 & 1 & 0.7186 & 0.7186 & 0.9999 & 0.9998 & 0.958 & 0.9999 & 0.9977 \\ \hline
Sadmind Amslverify Overflow & 0.7081 & 1 & 1 & 0.7138 & 0.7186 & 1 & 0.7081 & 0.7081 & 0.7055 & 0.4888 & 0.7274 & 0.7638 \\ \hline
FTP\_ Pass & 1 & 0.7013 & 0.7081 & 1 & 0.7186 & 0.7081 & 1 & 0.7081 & 0.7055 & 0.4888 & 0.7274 & 0.7638 \\ \hline
FTP\_ User & 1 & 0.7013 & 0.7082 & 1 & 0.9999 & 0.7081 & 0.7081 & 1 & 1 & 0.9621 & 0.9996 & 0.9966 \\ \hline
Rsh & 1 & 0.6986 & 0.7055 & 0.9999 & 0.9998 & 0.7055 & 0.7055 & 1 & 1 & 0.9632 & 0.9995 & 0.9963 \\ \hline
SSH\_ Detected & 0.9621 & 0.4804 & 0.4889 & 0.9599 & 0.958 & 0.4888 & 0.4888 & 0.9621 & 0.9632 & 1 & 0.9542 & 0.9365 \\ \hline
Telnet Terminal type & 0.9996 & 0.7274 & 0.7274 & 0.9998 & 0.9999 & 0.7274 & 0.7274 & 0.9996 & 0.9995 & 0.9542 & 1 & 0.981 \\ \hline
Mstream Zombie & 0.9966 & 0.7638 & 0.7638 & 0.9972 & 0.9977 & 0.7638 & 0.7638 & 0.9966 & 0.9963 & 0.9365 & 0.981 & 1 \\ \hline
\end{tabular}
\end{tiny}
\end{table*}

\begin{table*}[]
\caption{The correlation strength of alerts in 172.016.115.020 cluster using LLDOS 2.0.2}
\label{Table9}
\begin{scriptsize}
\begin{tabular}{|p{0.07\linewidth}|p{0.07\linewidth}|p{0.07\linewidth}|p{0.07\linewidth}|p{0.07\linewidth}|p{0.07\linewidth}|p{0.07\linewidth}|p{0.07\linewidth}|p{0.07\linewidth}|p{0.07\linewidth}|}
\hline
Alerts & Admind & Sadmind Amslverify Overflow & FTP\_ Pass & FTP\_ Put & FTP\_ User & Telnet Terminal type & Tenlnt envall & Telnet Xdisplay & Mstream Zombie \\ \hline
Admind & 1 & 1 & -0.9711 & -0.9711 & -0.9711 & -0.6975 & -0.6975 & -0.6975 & -0.5121 \\ \hline
Sadmind Amslverify Overflow & 1 & 1 & -0.9711 & -0.9711 & -0.9711 & -0.6977 & -0.6977 & -0.6977 & -0.5123 \\ \hline
FTP\_ Pass & -0.9711 & -0.9711 & 1 & 1 & 1 & 0.8485 & 0.8485 & 0.8485 & 0.7025 \\ \hline
FTP\_ Put & -0.9711 & -0.9711 & 1 & 1 & 1 & 0.8485 & 0.8485 & 0.8485 & 0.7025 \\ \hline
FTP\_ User & -0.9711 & -0.9711 & 1 & 1 & 1 & 0.8485 & 0.8485 & 0.8485 & 0.7025 \\ \hline
Telnet Terminal type & -0.6975 & -0.6977 & 0.8485 & 0.8485 & 0.8485 & 1 & 1 & 1 & 0.9727 \\ \hline
Tenlnt envall & -0.6975 & -0.6977 & 0.8485 & 0.8485 & 0.8485 & 1 & 1 & 1 & 0.9727 \\ \hline
Telnet Xdisplay & -0.6975 & -0.6977 & 0.8485 & 0.8485 & 0.8485 & 1 & 1 & 1 & 0.9727 \\ \hline
Mstream Zombie & -0.5121 & -0.5123 & 0.7025 & 0.7025 & 0.7025 & 0.9727 & 0.9727 & 0.9727 & 1 \\ \hline
\end{tabular}
\end{scriptsize}
\end{table*}

\begin{table*}[]
\caption{The correlation strength of alerts in 192.168.5.122 cluster using ISCX2012}
\label{Table10}
\begin{scriptsize}
\begin{tabular}{|p{0.12\linewidth}|p{0.12\linewidth}|p{0.15\linewidth}|p{0.1\linewidth}|p{0.17\linewidth}|p{0.1\linewidth}|}
\hline
Alerts & PROTOCOL-DNS potential dns & PROTOCOL-DNS TMG Firewall Client exploit attempt & ICMP & ET INFO SUSPICIOUS SMTP EXE ATTACHMENT & HTTP\_ INSPECT \\ \hline
PROTOCOL-DNS potential dns & 1 & 0.8484 & -0.9255 & -0.9977 & 0.2303 \\ \hline
PROTOCOL-DNS TMG Firewall Client exploit attempt & 0.8484 & 1 & -0.5847 & -0.811 & -0.3197 \\ \hline
ICMP & -0.9255 & -0.5847 & 1 & 0.9488 & -0.5818 \\ \hline
ET INFO SUSPICIOUS SMTP EXE ATTACHMENT & -0.9977 & -0.811 & 0.9488 & 1 & -0.2951 \\ \hline
HTTP\_ INSPECT & 0.2303 & -0.3197 & -0.5818 & -0.2951 & 1 \\ \hline
\end{tabular}
\end{scriptsize}
\end{table*}

\begin{table*}[]
\caption{The correlation strength of alerts in 192.168.2.107 cluster using ISCX2012}
\label{Table11}
\begin{scriptsize}
\begin{tabular}{|p{0.15\linewidth}|p{0.15\linewidth}|p{0.15\linewidth}|p{0.15\linewidth}|p{0.15\linewidth}|}
\hline
Alerts & ICMP test & ET POLICY PE EXE or DLL Windows file download HTTP & ET INFO Executable Retrieved With Minimal HTTP Headers & (spp\_frag3) Tiny fragment \\ \hline
ICMP test & 1 & 1 & -1 & -1 \\ \hline
ET POLICY PE EXE or DLL Windows file download HTTP & 1 & 1 & -1 & -1 \\ \hline
ET INFO Executable Retrieved With Minimal HTTP Headers & -1 & -1 & 1 &  \\ \hline
(spp\_frag3) Tiny fragment & -1 & -1 & 1 & 1 \\ \hline
\end{tabular}
\end{scriptsize}
\end{table*}

\section{Performance Evaluation}
\label{sec:11}

The performance of the proposed model is evaluated in terms of effectiveness. The effectiveness is measured by completeness and soundness of alert correlation. Completeness is computed as the ratio between the numbers of correctlycorrelated alerts by the number of related alerts (that belong to the same attack scenario). Soundness is defined as the ratio between the numbers of correctly correlated alerts by the number of correlated alerts. Rc and Rs are two simple measures designed to quantitatively evaluate completeness and soundness.

The experimental evaluation of the proposed model yielded datasets, soundness and completeness ranging between 80\% and 100 \% for the sample attack scenarios considered. By applying our model to DARPA 2000 (LLDOS1.0 and LLDOS 2.0.2 inside network) and ISCX 2012 datasets, five attack scenarios were constructed based on the compromised hosts (three of which were constructed from DARPA 2000 and two were constructed from ISCX 2012). As mentioned earlier, the attacker compromised five hosts in all datasets. Two hosts in LLDOS1.0 which are: locke (172.16.112.10) and pascal (172.16.112.50) network hosts and one host in LLDOS2.0.2 which is mill (172.16.115.20) host. While in ISCX 2012 dataset, two hosts were compromised which are 192.168.5.122 and 192.168.2.107 hosts. In each host, the attacker launches a different attack /step to reach their goal as shown in Figures 7, 8, 9, 10 and 11. 
Figure7 shows the scenario of the attack that compromised locke (172.16.112.10) from the inside network traffic in LLDOS1.0. There are five (5) alerts in this scenario, which are divided into five attack stages. The first stage contains Sadmind Ping alerts, which the attacker used to find out the vulnerable Sadmind services. The second stage comprises Sadmind Amslverify Overflow and Admind alerts. During this procedure, the attacker tried three different stack pointers and two commands in Sadmind Amslverify Overflow attacks for each victim host until one attempt succeeded as detailed in the description of the dataset. The third stage encompasses Rsh alerts, with which the attacker installed and started the mstream daemon and master programs. The fourth stage consists of alerts corresponding to the communications between the DDOS master and daemon programs. As can be seen clearly, Figure 7 represents the structure as well as the high level scenario of the sequence of the attacks which match the description of the attack scenario in the datasets and literature. The completeness of the attack scenario is measured at 80\% while the soundness is 100\%. Our further analysis indicates the reduction of the completeness due to un-correlated Rsh alerts with Mstream\_Zombie, which is one of the main steps of attack designed to complete their goal according to the description of data. 

\begin{figure}[h]
\includegraphics[width=0.6\textwidth]{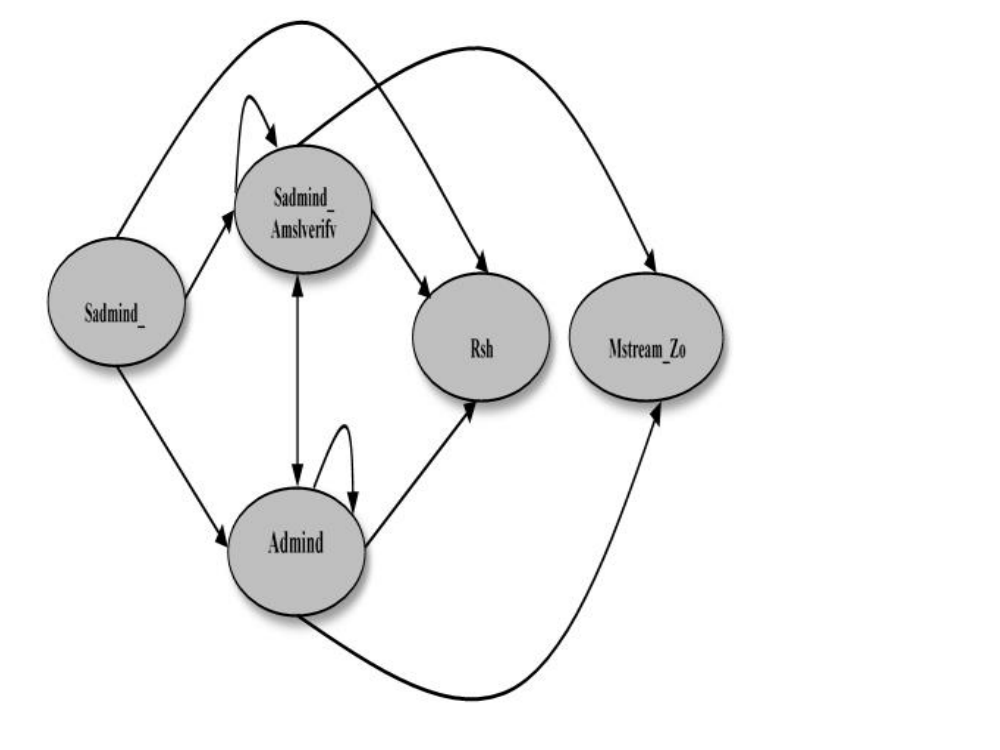}
\caption{\bf An attack scenario in 172.016.112.010 host in LLDOS 1.0}
\label{Fig7}
\end{figure}

\begin{figure}[h]
\includegraphics[width=0.6\textwidth]{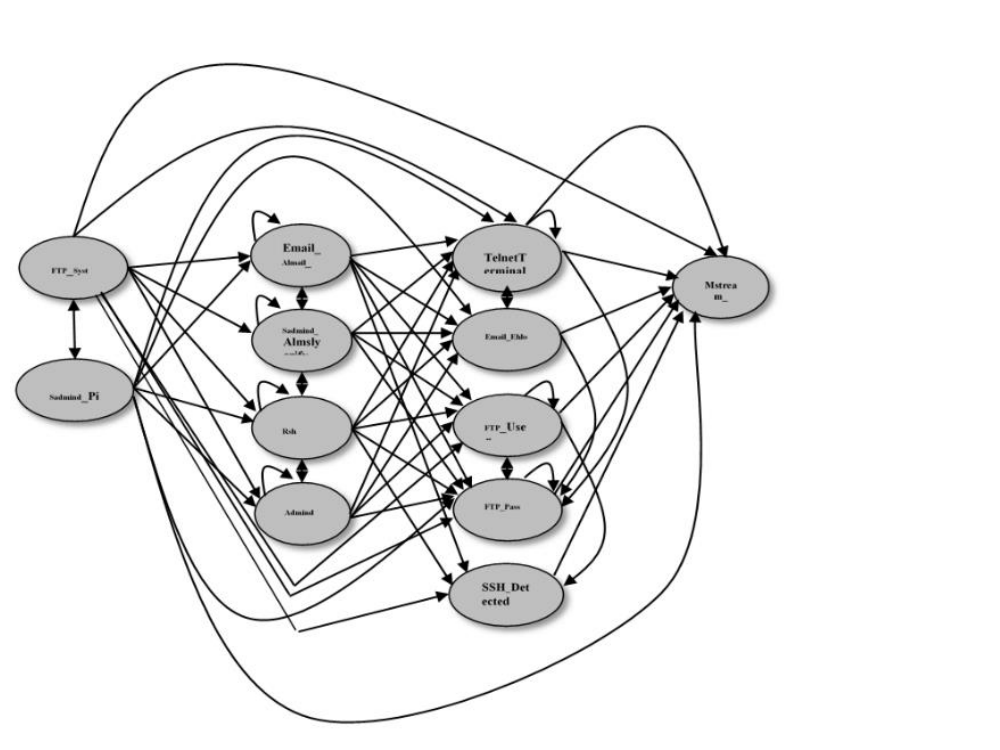}
\caption{\bf An attack scenario in 172.016.112.050 host in LLDOS 1.0}
\label{Fig8}
\end{figure}

\begin{figure}[h]
\includegraphics[width=0.6\textwidth]{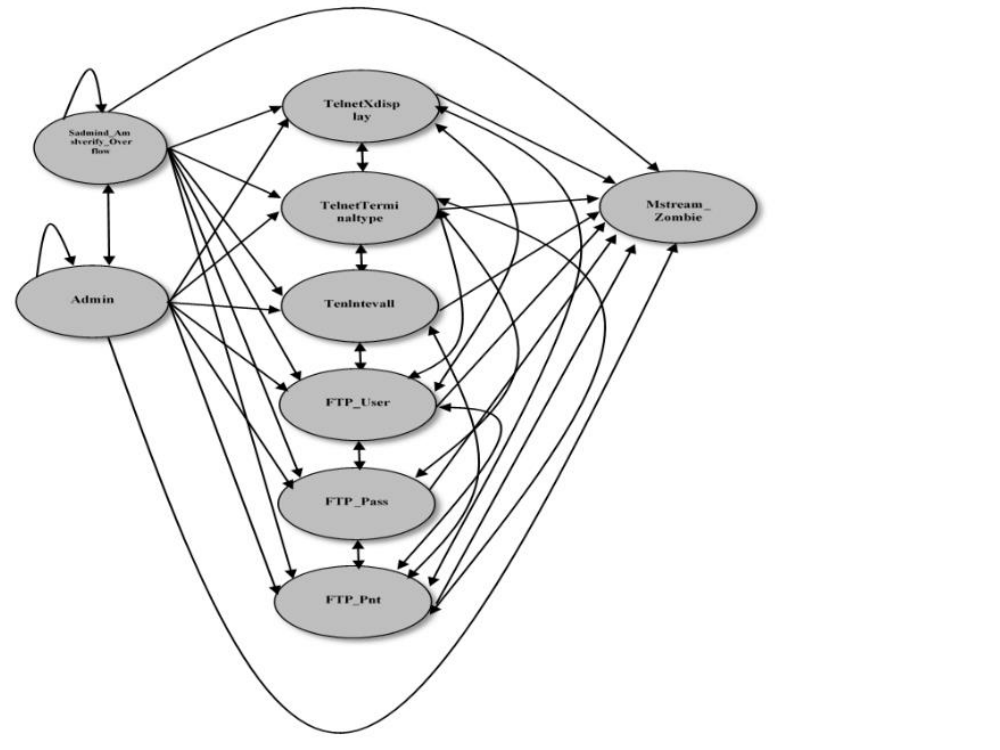}
\caption{\bf An attack scenario in 172.016.115.020 host in LLDOS 2.0.2}
\label{Fig9}
\end{figure}

\begin{figure}[h]
\includegraphics[width=0.6\textwidth]{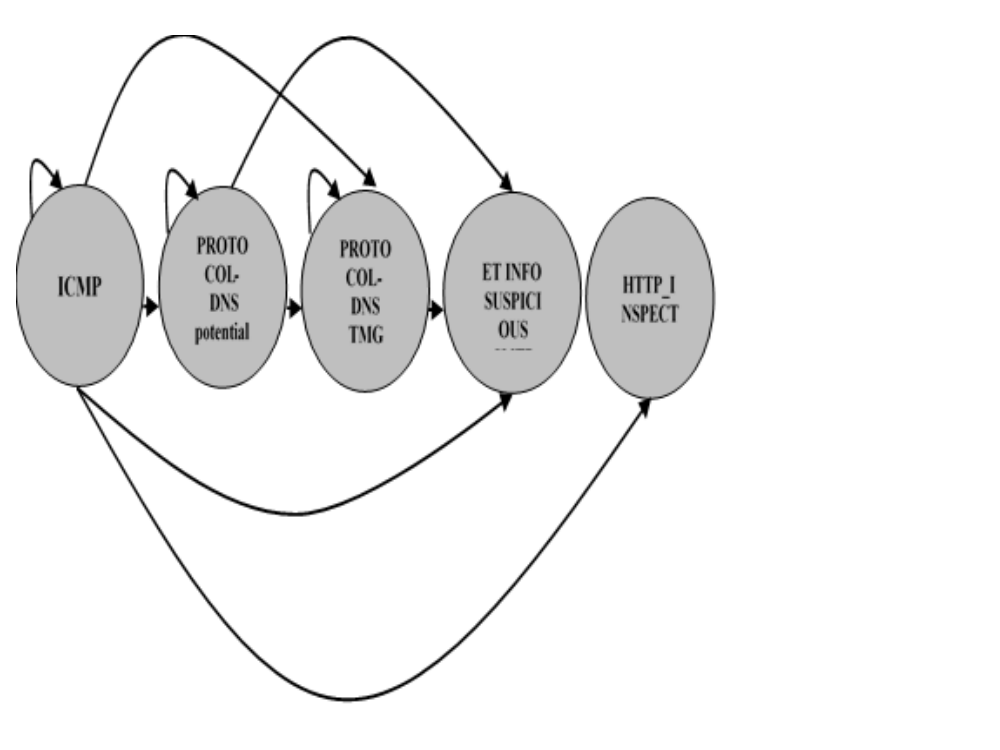}
\caption{\bf An attack scenario in 192.168.5.122 host in ISCX 2012}
\label{Fig10}
\end{figure}

\begin{figure}[h]
\includegraphics[width=0.6\textwidth]{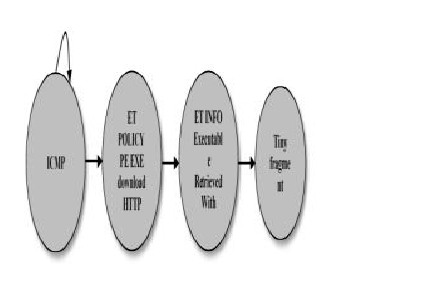}
\caption{\bf An attack scenario in 192.168.2.107 host in ISCX 2012}
\label{Fig11}
\end{figure}

Figure 8 presents the attack scenario that compromised the pascal (172.16.112.50) host. Unfortunately, there is no description of this attack scenario; neither in the datasets nor in the previous works.  Our further analysis showed that the attacker tried two ways to reach his goal. One of the ways used the same method as that described in the previous scenario. While in the second way, FTP\_Syst alert is triggered in the first stage during which the attacker issues a SYST command to File Transfer Protocol (FTP) server. The FTP server returns a response indicating the host operating system of the server knowing that the host operating system allows an attacker to customize an attack to exploit other vulnerabilities. The second stage consists of Email\_Almail\_Overflow and Email\_Ehlo. The attacker in this stage tries to affect Al Mail pop3 server by overflow of the buffer in the code that parses the SMTP headers or uses EHLO command to determine configuration on SMTP daemons, which trigger a considerable amount of Email\_Almail\_Overflow and Email\_Ehlo alerts. The third stage consists of some Telnets and FTP alerts, which the attacker used to access a victim host using environmental variables or file transfer protocol. The fourth stage consists of alerts corresponding to the communications between the DDOS master and daemon programs. As shown in the figure, all the related alerts are correlated together; based on the above analysis, it can be concluded that all correlated alerts are correctly correlated. Therefore, the completeness of the attack scenario in this scenario is 100\% while the soundness is also 100\%.

Figure 9 is the attack scenario extracted from LLDOS2.0.2. It is easy to find out that it has a similar pattern with the one extracted from LLDOS1.0. The attacker compromises several machines (in different ways) by exploiting the vulnerability of the Sadmind service and installs DDoS daemons on these machines by using either rsh or ftp. Realsecure does not raise alerts for the probing activity. Therefore, stage 1 is not shown in the corresponding attack scenario. The IDS identifies other stages by raising the following alerts: Sadmind Amslverify Overflow, Admind, FTP\_Put, FTP\_Pass, FTP\_User, TelnetXdisplay, TelnetTermnalType, Telntenvall, and MStream Zombie. All these alerts are correctly correlated. Thus, the completeness of the scenario is 100\% and the soundness is 100 \%. 
In ISCX 2012 two main attack scenarios based on 192.168.5.122and 192.168.2.107hosts were extracted successfully as shown in Figure 10 and 11. Figure 10 represents the scenario of the attack based on 192.168.5.122 infected host, the attack scenario starts when an attacker   applies an IP sweep attack to multiple hosts by sending an ICMP echo-request and listening for ICMP echo-replies to determine the active host. This activity triggers multiple ICMP alerts.  Then the attacker gathers information about the target host including network IP ranges, names servers, mail servers and user email accounts. This is achieved by querying the DNS for resource records using network administrative tools which causing a trigger for multiple PROTOCOL-DNS potential dns alerts. The IP sweep and DNS query are the main at-tack stages used by botnets which are mentioned in the first attack scenario (infiltrating the network from the inside). Therefore, the proposed model successfully constructs part of the actual first attack scenario. The others attacks in the first attack scenario (infiltrating the network from the inside) are missed in the proposed scenario because based on the ISCX documentation, these attacks were not detected by IDS in the ISCX network and as a result it cannot be reconstructed by our model.  Then, the Microsoft Fore-front Threat Management Gateway Firewall Client (TMG) vulnerability was used as a starting point. An attacker can exploit this issue to execute arbitrary code with the privileges of the user running the application. As a result of this stage large numbers of PROTOCOL-DNS TMG Firewall Client exploit attempt alerts are generated. After that an Internet Relay Chat (IRC) bot, is sent as an attachment for an update message. The bot has the ability to download malicious programs from remote servers and to execute them with user privileges. This activity triggers ET INFO SUSPICIOUS SMTP EXE- ATTACHMENT alerts. Finally, the bots were ordered to download the HTTP GET program using the download http get IP command. As each user finishes its download, a download complete message is sent over the channel. The Distributed Denial of Service attack is started using the run http IP threads command. As a result of this final stage large numbers of HTTP\_INSPECT alerts are generated.  Based on the documentation of the dataset observed that all the related alerts that represent the sequence of an attacker to achieve its goal are correctly represented in Figure 10. The complete relation among the related alerts in the figure is measured at 80\% while the soundness is 100\% . The reduction of the completeness due to uncorrelated ET INFO SUSPICIOUS SMTP EXE- ATTACHMENT alerts with HTTP\_INSPECT which is one of the main steps of attack designed to complete their goal according to the description of data. 
Figure 11 shows the scenario of the attack based on 192.168.2.107 infected host. This scenario describes the separate http denial of service attack. As can be seen clearly, Figure 11 represents the structure as well as the high-level scenario of the sequence of the attacks, which match the description of the attack scenario in the datasets and literature. Furthermore, all the related alerts in the figure have strong relation with other. Therefore, the completeness of the attack scenario in this scenario is 100 \% while the soundness is also 100\%. Table 12 shows the summary of the result regarding completeness and soundness of the five attack scenarios constructed in all datasets.

\section{CONCLUSION}
\label{sec:12}
This paper proposed an effective model for construction of an attack scenario. The construction of the attack scenario goes through three main processes, which are, identify related alert, mapping into relevance attack scenario and calculate correlation strength between alerts. Firstly, in identifying related alerts, alerts are grouped inter/intra stages based on the target IP address. While in the mapping process, hyper alert groups that do not contain complete attack stages are filtered out. Finally, the correlation strength between the alerts inside candidate groups that has complete attack stages is calculated to build the model .With this knowledge, the plan and the strategy of an attack can be determined and specified. The experimental results show that our proposed model produces an effective result. It offers complete coverage of possible correlations that alerts should have. One valuable avenue of later research is to combine all the information related to target attack recognition and risk assessment to develop a response plan.

\section*{Declaration}

This research has no Funding. 
\section*{Conflict of Interest }
 
The authors certify that they have no affiliation or interest in the subject matter or materials discussed in this manuscript. 

\section*{Availability of data and material}
The datasets generated during the study are available in the 
 (https://figshare.com/articles/DARPA\_2000\_dataset/4127157)
 (https://www.unb.ca/cic/datasets/)

\section* {Authors' Contributions}
All authors contributed to the study conception and design. Material preparation and analysis were performed by
[Taqwa Ahmed Alhaj] and [Mahyza MD Siraj]. The first draft of the manuscript was written by [Taqwa Ahmed Alhaj ] and all authors commented on previous versions of the manuscript. All authors read and approved the final manuscript.

\end{document}